\documentclass{emulateapj}
\usepackage{graphicx,amsmath,amssymb}
\usepackage{natbib}
\usepackage{apjfonts}
\bibpunct[, ]{(}{)}{;}{a}{}{,}

\newcommand{\sect}{\S\,}

\begin{document}

   \title{Very different X-ray to optical column density ratios in
          $\gamma$-ray burst afterglows: ionisation in GRB
          environments}

   \author{D.~Watson,\altaffilmark{1}
           J.~Hjorth,\altaffilmark{1}
	   J.~P.~U.~Fynbo,\altaffilmark{1}
           P.~Jakobsson,\altaffilmark{2}
           S.~Foley,\altaffilmark{3}
           J.~Sollerman,\altaffilmark{1,}\altaffilmark{4}
       and R.~A.~M.~J.~Wijers\altaffilmark{5}
            }
   \altaffiltext{1}{Dark Cosmology Centre, Niels Bohr Institute, University of Copenhagen, Juliane Maries Vej 30, DK-2100 Copenhagen \O, Denmark; darach, jens, jfynbo, jesper @dark-cosmology.dk}
   \altaffiltext{2}{Centre for Astrophysics Research, University of Hertfordshire, College Lane, Hatfield, Herts AL10 9AB, UK; P.Jakobsson@herts.ac.uk}
   \altaffiltext{3}{School of Physics, University College Dublin, Dublin 4, Ireland; sfoley@bermuda.ucd.ie}
   \altaffiltext{4}{Stockholm Observatory, Department of Astronomy, AlbaNova, S-106 91 Stockholm, Sweden}
   \altaffiltext{5}{Astronomical Institute ``Anton Pannekoek'', University of Amsterdam, Kruislaan 403, 1098 SJ Amsterdam, The Netherlands; rwijers@science.uva.nl}
   \begin{abstract}
     We compare the \ion{H}{1} column densities from Ly$\alpha$ absorption
     to the metal column densities from X-ray absorption in $\gamma$-ray
     burst (GRB) afterglows. Eight bursts of seventeen entering the sample,
     have significant extragalactic X-ray absorptions. Of these we find a
     range in metallicities from sub-solar to a few hundred times the solar
     value. There is a complete lack of correlation between the X-ray and
     optical column densities. This, and the large range and extreme values
     of these `metallicities', suggest that the column densities derived by
     one or both methods are not a reliable indication of the true total
     column densities towards GRBs. Ionisation of the GRB's gas cloud to
     large distances along the line of sight seems the most likely cause. 
     From the lower limit on the total column density and the UV luminosity
     of the GRBs, we derive a maximum distance to the majority of the gas
     surrounding GRBs of $\sim3$\,pc, suggesting that the gas probed by
     optical afterglow spectra is not the cloud in which
     the burst occurs. This is an encouraging result for the use of GRB
     optical afterglows as probes of the interstellar medium (ISM) in their
     host galaxies, as the ISM observed is less likely to be strongly
     affected by the GRB or its progenitor.
   \end{abstract}

   \maketitle

\section{Introduction\label{introduction}}

It is known from studies of neutral hydrogen absorption in $\gamma$-ray burst
(GRB) afterglows \citep[a technique pioneered half a decade
ago,][]{2001A&A...370..909J,2001A&A...373..796F,2003ApJ...597..699H}, that
they probe much higher column densities than QSO sightlines
\citep{2004A&A...419..927V}, almost certainly because GRBs probe the
star-forming environments in which they are born
\citep{2006A&A...460L..13J}. The highest column density GRB damped
Ly$\alpha$ absorbers (DLAs) are higher than any known QSO-DLA
\citep{2006A&A...460L..13J,2006ApJ...652.1011W,2005A&A...442L..21S,2005ApJ...634L..25C},
in spite of the much larger QSO-DLA sample.

Measurements of the \ion{H}{1} column density ($N($\ion{H}{1}$)$) are
needed to determine abundances in GRB environments where the metal column
densities are known either from absorption lines of non-refractory elements
such as Zn, or from the soft X-ray absorption which is dominated by
$\alpha$-chain elements in gas and solid phases. The optical
observations are complicated by the need to assume that certain elements are
not depleted by dust, and the need for high-resolution spectra. The X-ray
measurements do not suffer from these problems, but do not provide
a strong redshift constraint on the absorber
\citep[e.g.][]{2002A&A...395L..41W,2004ApJ...608..846S,2006A&A...455..803G}.
However using the X-ray metal column densities with Ly$\alpha$ columns has
been extremely difficult because of the conflicting redshift requirements
and has not yet been done in any object at $z>0.5$ other than a GRB. To
detect hydrogen Ly$\alpha$ 1216\,\AA\ with ground-based instruments we need
$z\gtrsim2$ to shift the line into the sensitivity range of optical/UV
spectrographs. But the soft X-ray absorption moves further and further out
of the bandpass as we go to higher redshift, such that once the neutral
oxygen edge at 0.52\,keV is no longer detectable, there is typically a
degeneracy between the redshift and the total absorption column that scales
roughly as $N_{\rm H}\propto (1+z)^{2.4}$. Thus we need a high
X-ray column density (greater than a few $\times10^{21}$) for it to
be detectable at $z\gtrsim2$.

With the advent of the \emph{Swift} satellite and its highly successful
X-ray telescope (XRT), very early, largely complete, and high flux
observations of GRB X-ray afterglows have been obtained for about ninety
GRBs every year. This means that the sample of GRB afterglow X-ray
absorption column densities has increased enormously, such that we now have
limits on the extragalactic X-ray absorption for almost every new GRB
discovered and we finally have a few GRB damped Ly$\alpha$ systems
(GRB-DLAs) where we can compare with significant detections of extragalactic
soft X-ray absorption: see GRB\,050401 \citep{2006ApJ...652.1011W},
GRB\,050505 \citep{2006MNRAS.368.1101H}, and GRB\,050730
\citep{2005A&A...442L..21S}.

In this \emph{Letter} we present the sample of GRB-DLAs with significant
extragalactic X-ray absorptions and compare the column densities obtained
with each method.
Uncertainties quoted are at the 68\% confidence level for one interesting
parameter unless otherwise stated. A cosmology where
$H_0=70$\,km\,s$^{-1}$\,Mpc$^{-1}$, $\Omega_\Lambda = 0.7$ and $\Omega_{\rm m}=0.3$ is assumed throughout.

\section{Sample and data reduction}\label{observations}

\emph{Swift} GRBs at $z>2$ were examined by \citet{2006A&A...460L..13J}
to determine the properties of GRB-DLAs. From that sample bursts with
well-constrained \ion{H}{1} columns were selected. We analysed XRT data from
these bursts where they were believed to have X-ray absorption in excess of
Galactic from the literature or the \emph{Swift} data
table.\footnote{\url[http://swift.gsfc.nasa.gov/docs/swift/archive/grb_table/]{http://swift.gsfc.nasa.gov/docs/swift/archive/grb\_table/}}
Analysis was done in a standard way. Spectra were fit with a power-law with
absorption fixed at the Galactic value and a second absorber at the host
redshift. Where the count rate was high, regions with the PSF-core excised
were used for spectral analysis to account for pile-up.

To ensure that only statistically significant excess absorption columns were
selected, two conservative criteria were applied. 1) GRBs at low Galactic
latitude ($|b|<20\degr$) were excluded. 2) The excess column density had to
be detected above the 99.7\% confidence level when the $z=0$ absorber was
fixed at $1\times10^{20}$cm$^{-2}$ or 20\% above the Galactic value,
whichever was larger
\citep[see][]{1986ApJ...310..291E,1989AJ.....97..777E,1990ARA&A..28..215D}.
Galactic column densities obtained from the newer \ion{H}{1} survey of
\citet{2005A&A...440..775K} were consistent with those obtained from
\citet{1990ARA&A..28..215D} within $\sim3\times10^{19}$\,cm$^{-2}$. At early
times an absorbed power-law model fit occasionally yields excess absorption
not present at later times
\citep[e.g.][]{2005A&A...442L..21S,2007A&A...462..565G,2007ApJ...654L..17C}.
\citet{2006astro.ph.12564B} have examined the reasons behind this and we
discuss it in \sect\ref{discussion}. Because of this effect we adopt the
lower value of absorption where inconsistent values were found at early and
later times. GRBs\,050730, 050820A, and 060714 are the bursts
affected by this, and we have adopted absorption values from the later PC
mode spectra for these bursts.

The spectra of GRBs\,050401, 050505, 050730 and 050904 have been presented
elsewhere
\citep{2006ApJ...652.1011W,2006MNRAS.365.1031D,2006MNRAS.368.1101H,2005A&A...442L..21S,2006ApJ...637L..69W,2007A&A...462..565G,2007ApJ...654L..17C}.
For the bursts with persistent excess absorptions published
previously---050319 \citep{2006A&A...449...61C}, 050401, 050505, and
050904---we obtain somewhat different values. This may be due to the more
recent calibration used in this analysis. However, in each case, the
authors claim a detection of excess absorption and adopting those values do
not change our conclusions.

\section{Results}\label{results}

\begin{figure}
 \includegraphics[angle=-90,bb=63 54 559 386,clip=,width=\columnwidth]{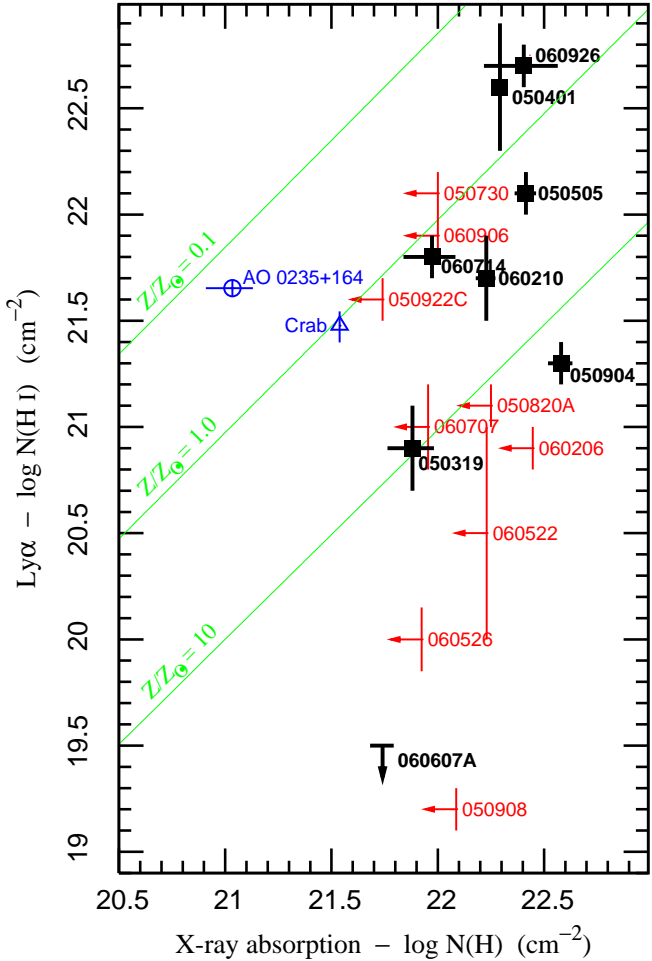}
 \caption{Neutral hydrogen column density (from the UV) as a function of
          X-ray equivalent hydrogen column density. The ratio of these
          values provides an estimate of the metal abundance of the
          absorbing medium. The clear lack of correlation between these
          column densities seems initially to indicate an extremely strong
          variation in the metal enrichment of the absorbing clouds near
          GRBs. These values are so extreme however, and many so much
          greater than solar, that it seems likely that there are strong
          systematic uncertainties related to either or both of the
          absorbing column density estimates from X-rays and the UV,
          possibly because of ionisation of most of the line-of-sight
          hydrogen. Data from absorption toward a Galactic source (the Crab
          pulsar) and a high column density DLA (foreground to the blazar
          AO\,0235+164) are shown for comparison. Where the uncertainties on
          the \ion{H}{1} column is not in the literature a value of 0.1 has
          been plotted.}
 \label{fig:xrayoptical}
\end{figure}

Seventeen GRBs enter our sample as set out in Table~\ref{tab:columns}. The comparison of
X-ray and and hydrogen Ly$\alpha$ absorption is shown in
Fig.~\ref{fig:xrayoptical}. If the metallicities of GRB formation sites are
low
\citep{2003A&A...400..499L,2003A&A...406L..63F,2006Natur.441..463F,2007astro.ph..1246M,2007AJ....133..882K},
we would expect to make no significant detections of excess absorption in
this sample. Instead, eight GRBs have significant excess
column densities: GRBs\,050319, 050401, 050505, 050904, 060210, 060607A,
060714, and 060926 (see Table~\ref{tab:columns} and
Fig.~\ref{fig:xrayoptical}).

\begin{table}
\caption{UV and X-ray estimates of the absorbing column densities in GRBs}
\label{tab:columns}
\setlength{\tabcolsep}{6pt}
 \begin{center}
  \begin{tabular}{@{}lccccc@{}}
   \hline\hline
   GRB	& $z^a$	& $\log{N_{\rm H\,{\sc I}}}^a$	& Total $N_{\rm H}$	& Gal.\ $N_{\rm H}$	& $N_{\rm H}$ at $z$	\\
	&	&				& at $z=0$		& 	&			\\
	& 	& 				& \multicolumn{3}{c}{($10^{20}$\,cm$^{-2}$)}			\\
   \hline
050319		& 3.24	& $20.9$	& $5.1^{1.1}_{-1.0}$	& $1.1$		& $76^{+20}_{-18}$	\\ %1 %%Numbers from us. Campana et al. published slightly lower values
050401		& 2.90	& $22.6$	& $16.9\pm0.8$		& $4.8$		& $196\pm13$ 		\\ %2 %%Total nh at z=0 estimated from the other values in our paper. Also estimated from simulations and gave same results.
050505		& 4.27	& $22.1$	& $9.8\pm1.0$		& $2.0$		& $130\pm36$		\\ %3 %%Got data for 050505. !!REMEMBERED PILE-UP!! No WT!?
050730		& 3.97	& $22.1$	& $3$			& $3.05$	& $<100$		\\ %5 %%Got from Rhaana's paper.
050820A 	& 2.61	& $21.1$	& $4.0\pm1.0$		& $4.71$	& $<178$ 		\\ %6
050904		& 6.30	& $21.3$	& $9.1\pm0.6$		& $4.9$		& $380\pm50$	 	\\ %7
050908		& 3.34	& $19.2$	& $2.2\pm1.5$		& $2.1$		& $<122$		\\ %8 %%Very little data. Dominated by flares. No excess.
050922C		& 2.20	& $21.6$	& $6.5^b$		& $5.75$	& $<55^c$		\\ %9 %%*From Swift Table
060206		& 4.05	& $20.9$	& $5.2^{+1.4}_{-1.3}$	& $0.9$		& $<280$		\\ %13 %%Did PC data for 060206 (no pileup correction done for PC). (Late starting observation?)
060210		& 3.91	& $21.7$	& $15.2\pm0.5$		& $8.5$		& $169\pm18$		\\ %14 %%!!REMEMBERED PILE-UP!! Very close to b=20!
060522		& 5.11	& $20.5$	& $4.6^b$		& $4.83$	& $170^c$		\\ %17 %%*From Swift Table
060526		& 3.22	& $20.0$	& $6^b$			& $5.51$	& $<84^c$		\\ %18 %%*From Swift Table
060607A		& 3.08	& $<19.5$	& $5.6\pm0.4$		& $2.7$		& $55\pm7$	 	\\ %20
060707		& 3.43	& $21.0$	& $1.8^b$		& $1.76$	& $<90^c$		\\ %21 %%*From Swift Table
060714		& 2.71	& $21.8$	& $13.6^{+2.0}_{-1.9}$	& $6.7$		& $94^{+27}_{-25}$	\\ %22 %%NO pile-up problem... but variations in nH
060906		& 3.69	& $21.9$	& $9.66^b$		& $9.66$	& $<100^c$		\\ %23 %%*From Swift Table
060926		& 3.21	& $22.7$	& $20.5^{+4.8}_{-4.2}$	& $7.31$	& $250^{+110}_{-90}$	\\%25 %%
   \hline
  \end{tabular}
 \end{center}
$^a$ See \citet{2006A&A...460L..13J} and references therein.

$^b$ From \url[http://swift.gsfc.nasa.gov/docs/swift/archive/grb_table/]{http://swift.gsfc.nasa.gov/docs/swift/archive/grb\_table/}

$^c$ Estimated using a redshift-corrected, $3\times10^{20}$ uncertainty in the
    X-ray column and a $1\times10^{20}$ uncertainty in the Galactic column
    density. 
\end{table}

The best limits that can be placed on the column density from
\emph{Swift}-XRT data above $z=2$ is roughly $\log{N_{\rm H}}\lesssim 21.5$.
Fig.~\ref{fig:xrayoptical} is therefore not populated below 21.5 in X-ray column.
Such a limit could only constrain the metal abundance to be less than solar
in the highest \ion{H}{1} column density cases.
Upper limits obtained in the optical/UV are more interesting: GRB\,060607A
has a tightly constrained, low \ion{H}{1} column density, but large
X-ray absorption. (The same is true for GRB\,060124,
but it does not enter our sample at $b=17\degr$.)

There is evidence of a spread in the metallicities of GRB hosts, though
with most fairly close to $\sim10$\% of the solar value \citep[see][]{2006A&A...451L..47F}. If the metallicities of GRB
environments were all approximately similar we would expect some correlation
between the optical and X-ray column density determinations. It is clear
from Fig.~\ref{fig:xrayoptical} that there is no such correlation and that
the optical-to-X-ray ratios span a range of at least a few hundred.
Furthermore, the metal abundances derived by this comparison indicate
metallicities far above the solar values. The extreme abundances and the
lack of correlation between the optical and X-ray values are disturbing and
strongly imply that the X-ray absorption and the Ly$\alpha$ absorption are
sampling different environments. These data immediately show that for GRBs
the X-ray-to-optical absorption ratio is not a useful measure of the
metallicity.

\subsection{Reliability of the measurements}
Metal abundances have been determined for intervening extragalactic
absorbers in a few radio-loud AGN using X-ray and Ly$\alpha$ absorption.
Those values are significantly below solar
\citep[see][]{2003ApJ...590..730T}. The metallicities are consistent with
extinction measures and expectations for these objects.
For the Crab
pulsar, this technique yields values consistent with solar
metallicity \citep{2000ApJ...537..861S,2001A&A...365L.212W}. Both these
results are plotted for comparison on Fig.~\ref{fig:xrayoptical} and
indicate that in general the comparison of column densities from X-ray
and Ly$\alpha$ is valid for both Galactic and extragalactic sources.

GRB afterglows exhibit curvature close to the cooling frequency, which may
be at X-ray wavelengths at certain times and may mimic absorption.
\citet{2006astro.ph.12564B} have shown that in many cases where the soft
X-ray absorption apparently changes rapidly in the first few hundred second
after the burst, this may be better explained by curvature of the
continuum than ionisation of the absorber. This seems a reasonable
explanation in the early phases of some bursts. Such early curvature
however, should not affect the results presented here, as we have used
absorption values from later spectra where early (high) values of the
absorption are inconsistent with later data. Such an approach is
conservative, in the sense that it leads to lower column
densities. In general, spectral curvature is not responsible for the
detection of excess absorption. Most of the bursts in question have early WT
and PC mode data that agree on the absorption to within 1$\sigma$
despite large spectral changes.

Given that the technique is reliable, could it be that the Galactic column
densities are underestimated? This explanation can be ruled out. First of
all we have excluded any excess column density that is not significant above
the 90\% variations of the Galactic columns
\citep{1986ApJ...310..291E,1989AJ.....97..777E,1990ARA&A..28..215D}.
Furthermore solar abundances were used to convert the Galactic 21\,cm
column densities \citep{1990ARA&A..28..215D}, resulting in a high
(conservative) estimate of the foreground Galactic column, higher
than is often argued for the Galactic gas by e.g.\
\citet{2000ApJ...542..914W}. A large underestimate of the Galactic
column density can also be excluded on the basis of observations of blazars with \emph{BeppoSAX}
\citep{2005A&A...433.1163D} that show no substantial absorption compared to
the values of \citet{1990ARA&A..28..215D}. Of nearly 90 blazars examined
with \emph{BeppoSAX} only 17 showed excess absorption in the
observed 0.1--50\,keV spectra and even these excesses are small,
$\sim1\times10^{20}$\,cm$^{-2}$. 
Finally, if the excess columns were due to Galactic foreground
absorption, we would expect a systematic rise in observed column
densities with redshift, an effect we do not observe when we combine with
the lower-redshift values of \citet{2006A&A...449...61C}.

\section{Discussion}\label{discussion}

It is only with the success of \emph{Swift} that we have been
able to obtain Ly$\alpha$ and X-ray absorption measures in the same bursts.
Given the expectations for neutral hydrogen column densities surrounding
GRBs \citep{2002ApJ...565..174R}, few significant detections of excess
absorption in the X-ray afterglows of GRBs should have been expected above
$z\gtrsim2$.

\subsection{Intervening absorbers}

It is expected that high redshift sources \citep[and GRBs have a very high
mean redshift,][]{2006A&A...447..897J} will often have intervening line of
sight absorbers. The X-ray absorption does not tell us about the redshift of
the absorber and some absorption could in principle be related to low-$z$
systems. The observed absorption in X-rays drops substantially as the
absorber is moved to higher redshift, so for this to be an effective
explanation any absorber would have to be at relatively low redshift.
However, very large column densities of metals would routinely be required
along most lines of sight at low redshifts. Recently,
\citet{2006ApJ...648L..93P} have shown that GRBs have intervening large
column density (W$_\lambda>1$\,\AA) \ion{Mg}{2} absorbers about four times
more frequently than QSOs. And it is a suggestive coincidence that GRBs also
have significant excess soft X-ray absorptions. However, to explain the
results observed here with intervening absorbers is not possible. This would
require an absorber with typically $\log(N_{\rm H})\simeq22.1$ at $z=0.3$
(assuming $Z/Z_\odot=0.1$) in half of the GRB sightlines---an extremely
large value. Assuming that an \ion{Mg}{2} absorber with
W$_\lambda\sim1-3$\,\AA\ corresponds to $\log(N_{\rm H})\lesssim21$ at this
metallicity \citep{2007MNRAS.tmp..112M}, the required absorption is roughly
two orders of magnitude above that provided by the \ion{Mg}{2} absorbers
seen in GRBs at $z\lesssim0.5$ \citep{2006ApJ...648L..93P}. To reinforce
this point, a similar excess of large equivalent width \ion{Mg}{2} absorbers
is also found in the foregrounds of blazars \citep{1997ApJ...489L..17S},
which, as noted above, do not have large X-ray absorptions
\citep{2005A&A...433.1163D}. This is an important difference in that it
seems likely that GRBs and blazars sample similar column densities of
intervening absorbers, but GRBs typically have large soft X-ray absorptions
and blazars do not,
showing that the large absorptions are indeed intrinsic
to the GRB hosts.

It has been suggested \citep{2006A&A...460L..13J} that the lack of very high
column density and apparent overabundance of low column density
sources in comparison to model predictions may be due to ionisation of the
hydrogen near the GRB. This hypothesis explains the results found here; the
hydrogen may be ionised to large distances by the early afterglow, but
the metals, while also ionised, will not be stripped to such
an extent that they cease substantially to absorb the soft X-rays. This
hypothesis is also consistent with the lack of expected Wolf-Rayet features
in most GRB optical afterglows \citep{2006astro.ph.11079C}.

\subsection{Size of the gas cloud}

It is possible to limit the size of the gas cloud where most of the hydrogen
has been ionised by the GRB. A maximum distance to the gas is found by
assuming the X-ray column density represents the total column density
(it cannot obviously be smaller than this, and if it is larger, then the
ionisation radius will be smaller) and that each ionising UV photon is
intercepted by a hydrogen atom (if this is not the case, the radius will
again, necessarily be smaller). Since we know the total UV fluence for
most of these bursts is a small fraction of the prompt energy, we can
derive a maximum distance to which the GRB could have ionised the gas. In
all cases, this is less than $\sim3$\,pc. Interestingly, if the gas cloud is
already ionised by the massive stars in the region, the typical densities of
\ion{H}{2} regions ($\lesssim10^4$\,cm$^{-3}$) yield sizes which are also at
most a few parsecs in radius. Smaller radii require higher densities. Radii
an order of magnitude greater than this make the total masses unfeasibly
large. This radius is similar to that obtained by
\citet{2007ApJ...654L..17C} based on the assumption that the apparent
decreasing absorption detected in GRB\,050904 is due to ionisation by the
GRB.

Observations of UV absorption lines have allowed distances to low-ionisation
gas to be derived that are far larger than the maximum distance to the
majority of the gas column we derive above. \citet{2006ApJ...648...95P} find
$\gtrsim50$\,pc for \ion{Mg}{1} in GRB\,050111, and
\citet{2006astro.ph.11690V} find $1.7\pm0.2$\,kpc for \ion{Ni}{2} and
\ion{Fe}{2} in GRB\,060418. Gas with high ionisation states closer to the
GRB has been detected in the optical \citep{2006astro.ph..9825D}. Though
1.7\,kpc seems large to be the typical distance scale between clouds in most
bursts, taken together, those results support our conclusion in this paper,
that the optical/UV spectra are probing gas that is outside the pc-scale
structure in which the GRB explodes.

\section{Conclusions\label{conclusions}}

We have examined the X-ray afterglow spectra of the sample of \emph{Swift}
GRBs at $z>2$ as defined by \citet{2006A&A...460L..13J} and compared their
extragalactic soft X-ray absorbing column densities to the host \ion{H}{1}
column densities. We find no correlation between the column density values,
and the X-ray absorptions often far exceed the \ion{H}{1} column densities.
The most likely explanation for this discrepancy is that GRBs substantially
ionise the hydrogen in their immediate environments (at least along the line
of sight). This may be consistent with the sublimation of the dust that is
expected (by analogy with the local universe) to be associated with large
X-ray column densities, though it does not provide evidence in favour of
dust sublimation. It seems likely that significant information about the
immediate circumburst environment may be gleaned from high-resolution X-ray
spectroscopy of the very early afterglow, probably revealing the complex
absorption pattern associated with metal-rich warm absorbers, as observed in
some AGN. Finally, we find the maximum distance for the bulk of the
absorbing material surrounding the GRBs to be $\sim3$\,pc for these bursts,
similar to the sizes of compact \ion{H}{2} regions. Since the distances
inferred from optical observations are typically larger than this, it
implies that the optical/UV spectroscopy of GRB afterglows typically probes
environments that are little affected by the progenitor, the explosion or
the bright afterglow. This is particularly encouraging for the use of GRBs
as probes of star-forming regions at high redshift.

\begin{acknowledgements}
We thank Rhaana Starling, Tamara Davis, Nat Butler and Enrico Ramirez-Ruiz
for useful discussions. The Dark Cosmology Centre is funded by the DNRF.
\end{acknowledgements}

\end{document}